\documentclass[fleqn,usenatbib]{mnras}

\usepackage{amsmath,graphicx,amssymb}
\usepackage{txfonts}
\usepackage{natbib}
\usepackage{subfigure}
\usepackage{hyperref}
\usepackage{adjustbox}

\defcitealias{lb16}{LB16}

\title[V680 Mon $-$ a young HgMn star in an EB heartbeat system]{V680 Mon $-$ a young mercury-manganese star in an eclipsing heartbeat system}

\author[Paunzen et al.]{
Ernst Paunzen$^{1}$\thanks{E-mail: epaunzen@physics.muni.cz}
Stefan H{\"u}mmerich,$^{2,3}$
Miroslav Fedurco,$^{4}$
Klaus Bernhard,$^{2,3}$
\newauthor
Richard Kom\v{z}\'{\i}k,$^{5}$
Martin Va\v{n}ko$^{5}$
\\
$^{1}$Department of Theoretical Physics and Astrophysics, Masaryk University, Kotl\'a\v{r}sk\'a 2, 611 37 Brno, Czech Republic\\
$^{2}$Bundesdeutsche Arbeitsgemeinschaft f{\"u}r Ver{\"a}nderliche Sterne e.V. (BAV), Berlin, Germany\\
$^{3}$American Association of Variable Star Observers (AAVSO), Cambridge, USA\\
$^{4}$Institute of Physics, Faculty of Science, P. J. \v{S}af\'{a}rik University, Ko\v{s}ice, Slovak Republic\\
$^{5}$Astronomical Institute of the Slovak Academy of Sciences, 
059 60 Tatransk\'{a} Lomnica, Slovak Republic\\
}

\date{Accepted XXX. Received YYY; in original form ZZZ}

\pubyear{2021}

\begin{document}

\label{firstpage}
\pagerange{\pageref{firstpage}--\pageref{lastpage}}
\maketitle

\begin{abstract}\
Chemically peculiar stars in eclipsing binary systems are rare objects that allow the derivation of fundamental stellar parameters and important information on evolutionary status and the origin of the observed chemical peculiarities. Here we present an investigation of the known eclipsing binary system BD+09 1467 = V680 Mon. Using spectra from the Large Sky Area Multi-Object Fiber Spectroscopic Telescope (LAMOST) and own observations, we identify the primary component of the system as a mercury-manganese (HgMn/CP3) star (spectral type kB9 hB8 HeB9 V HgMn). Furthermore, photometric time series data from the Transiting Exoplanet Survey Satellite (TESS) indicate that the system is a 'heartbeat star', a rare class of eccentric binary stars with short-period orbits that exhibit a characteristic signature near the time of periastron in their light curves due to the tidal distortion of the components. Using all available photometric observations, we present an updated ephemeris and binary system parameters as derived from a modelling of the system with the ELISa code, which indicate that the secondary star has an effective temperature of $T_{eff}$\,=\,$8300_{-200}^{+200}$ (spectral type $\sim$A4). V680 Mon is only the fifth known eclipsing CP3 star, and the first one in a heartbeat binary. Furthermore, our results indicate that the star is located on the zero-age main sequence and a possible member of the open cluster NGC\,2264. As such, it lends itself perfectly for detailed studies and may turn out to be a keystone in the understanding of the development of CP3 star peculiarities.
\end{abstract}

\section{Introduction} \label{sec:intro}

Among the zoo of chemically peculiar (CP) stars, the mercury-manganese (HgMn/CP3) stars form a rather homogeneous group. They are traditionally identified by the presence of strong \ion{Hg}{ii} and \ion{Mn}{ii} lines in optical spectra and occupy the rather restricted spectral-type range between B6 and A0 \citep{preston74,smith96,chojnowski20,paunzen20}. In addition to strong atmospheric overabundances of Hg and Mn (up to 6 and 3 dex over Solar, respectively; e.g. \citealt{white76}, \citealt{smith96}, \citealt{ghazaryan16}), CP3 stars exhibit numerous other peculiarities, in particular a general overabundance of heavy elements. Generally, the strength of the overabundances increases with atomic number \citep{castelli04,ghazaryan16}. Detailed information on the chemical composition of CP3 stars has for example been provided by \citet{castelli04} and \citet{ghazaryan16}.

CP3 stars are slow rotators \citep{mathys04} and have a high rate of multiplicity. Multiplicity frequencies have been estimated at more than 50\,\% \citep{smith96}, with values up to 67\,\% \citep{hubrig95} and 91\,\% \citep{schoeller10}. CP3 stars do not show the strong magnetic fields that are observed in the Ap/CP2 and the He-peculiar stars (which are lumped together under the label 'magnetic CP stars') and are generally listed with the non-magnetic CP stars. However, several recent studies announced the presence of weak or tangled fields \citep{hubrig10,hubrig12} and this has remained a controversial issue \citep{kochukhov13,hubrig20}. CP3 stars show an inhomogeneous surface element distribution ('chemical spots') with obvious signs of secular evolution (e.g. \citealt{hubrig95}, \citealt{adelman02}, \citealt{hubrig06}, \citealt{kochukhov07}, \citealt{briquet10}, \citealt{korhonen13}).

CP3 stars are relatively rare objects. However, recent progress has led to a substantial extension of the number of known CP3 stars \citep{chojnowski20,paunzen20}. At the time of this writing, more than 550 Galactic CP3 stars have been registered. Furthermore, with the advent of space photometry, an increasing number of CP3 stars has been found to be photometric variables \citep{alecian09,balona11,morel14,paunzen13,strassmeier17,white17,huemmerich18}. Generally, current studies favoured rotational over pulsational modulation as the cause of the observed variability in the investigated stars (\citealt{huemmerich18}, and the discussion therein).

Given their high multiplicity rate, the presence of a CP3 star component in a number of eclipsing binary systems would be expected. These objects, however, are exceedingly rare: the present authors are only aware of four eclipsing binaries containing a CP3 star component, viz. HD 34364 = AR Aur \citep{hubrig06,folsom10}, HD 161701 = 	HR 6620 \citep{gonzalez14}, TYC 455-791-1 = HSS 348 \citep{strassmeier17}, and HD 10260 = V772 Cas \citep{kochukhov20}. 

As binaries -- in particular eclipsing ones -- allow the derivation of fundamental stellar parameters like mass and radius, and only very few CP stars have direct determinations of these parameters \citep{north04}, the discovery of binary systems with CP star components is important. Furthermore, such systems may help to understand, and put constraints on, the processes responsible for the formation of the observed chemical peculiarities and the time scales involved.

Here we report on the discovery of another eclipsing binary system containing a CP3 star component, viz. BD+09 1467 = V680 Mon, which is well suited to follow-up studies dealing with the solution of the system and the determination of exact stellar parameters for both components. We show that V680 Mon is a 'heartbeat' star, a rare class of eccentric binary stars with short-period orbits (1\,d\,$\la$\,$P_{orb}$\,$\la$\,1\,yr) that exhibit a characteristic signature near the time of periastron in their light curves whose shape is reminiscent of an electrocardiogram diagram (hence the name).

Section \ref{sec:analysis} provides information on our target star, its astrophysical parameters and location in the sky, and the observations. We present our results in Section \ref{sec:results} and discuss them in Section \ref{sec:discussion}.

\begin{figure}
        \includegraphics[trim = 0mm 0mm 18mm 105mm, clip, width=\columnwidth]{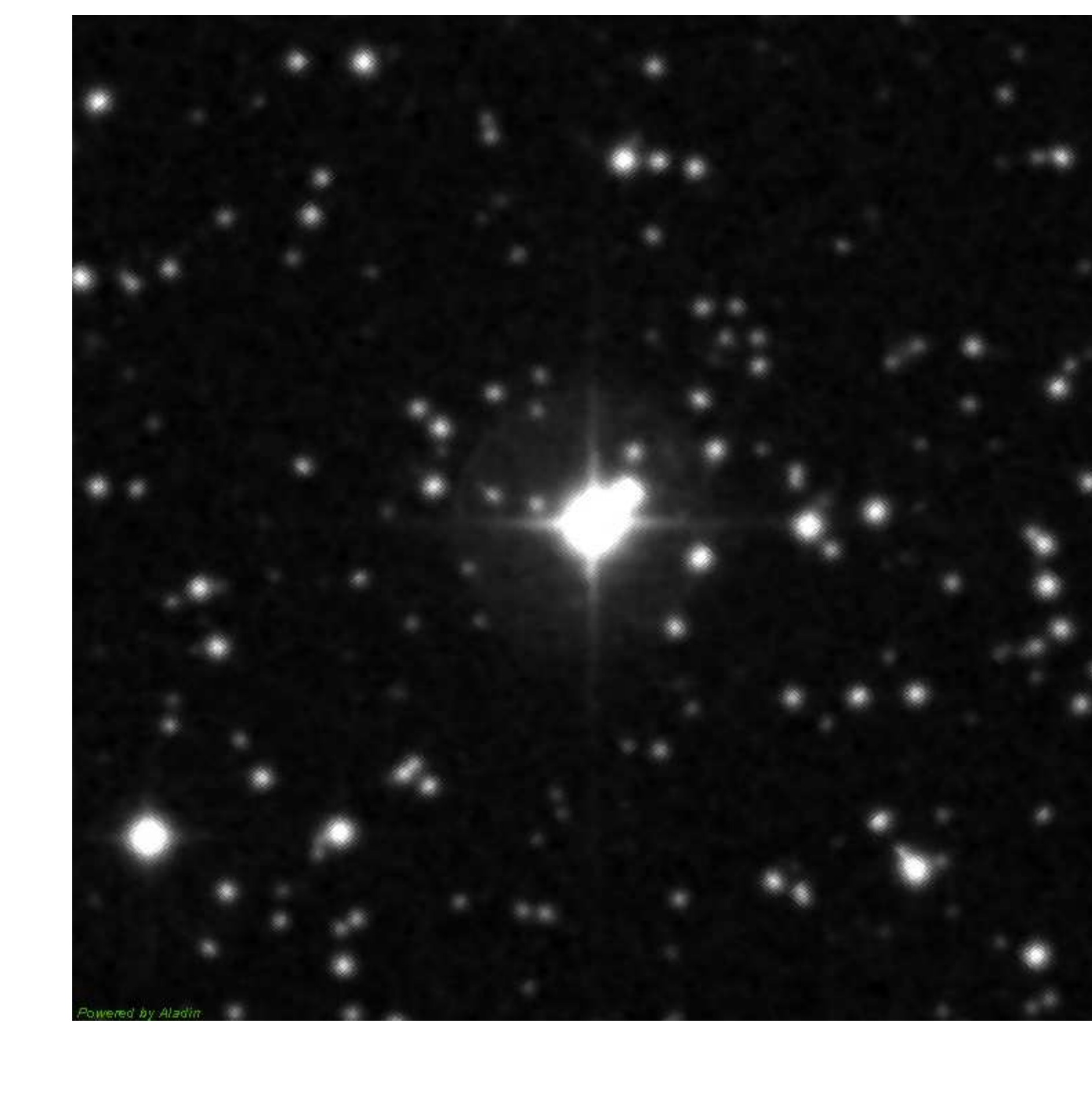}
    \caption{Sky region of V680 Mon on DSS2 blue plates, accessed via the ALADIN sky atlas \citep{ALADIN1,ALADIN2}. The field of view is 4.289\arcmin\,x\,4.108\arcmin. North is at the top, east is at the left. The image is centered on V680 Mon. The bright star to the north-west that appears partially blended with V680 Mon is 2MASS J06593015+0919070.}
    \label{skyview}
\end{figure}

\begin{table*}
\caption{Basic stellar parameters of V680 Mon (2MASS J06593071+0918596) and its close companion 2MASS J06593015+0919070.}
\label{table_companion}
\begin{adjustbox}{max width=1.0\textwidth}
\begin{tabular}{llllllllllll}
\hline
\hline
ID & Gaia DR2 & $\alpha$ & $\delta$ & $\pi$ & e\_$\pi$ & $G$\,mag & e\_$G$\,mag & $(BP-RP)_0$ & e\_$(BP-RP)_0$ & MV$_0$	& e\_MV$_0$ \\
\hline
J06593071+0918596	& 3157882748862195072 & 104.8779802	& 9.3165762	& 1.594 & 0.146 & 9.9803	& 0.0011 & 0.000	& 0.004 & 1.09 &	0.16 \\
J06593015+0919070 & 3157882744564424704 & 104.8756241	& 9.3186280 & 0.370	& 0.018 & 12.5163	& 0.0002 & 1.146	& 0.002 & 0.35 & 0.09 \\
\hline
\hline
\end{tabular}                          
\end{adjustbox}
\end{table*}

\section{Target star, observations and data analysis} \label{sec:analysis}

\subsection{Target star} \label{subsec:target_star}

V680 Mon = BD+09 1467 = HD 267564 (spectral type B8, \citealt{cannon93}; $V$\,=\,10.13\,mag, \citealt{HIPPARCOS}; $G$\,=\,9.98\,mag, \citealt{gaia2}) was identified as a variable star by \citet{parenago46}, who listed it under the preliminary designation of SVS 1025 Monocerotis and suggested it to be an eclipsing binary with a range of 9.5\,$-$\,10.1 mag ($pg$). No period could be derived from the available observations. The star was included as NSV 3233 into the New Catalogue of Suspected Variable Stars \citep{NSV} and later entered the General Catalogue of Variable Stars as V680 Mon \citep{kholopov87,GCVS}.

V680 Mon has been little studied, and discrepant information is found in the literature. From an analysis of 85 photographic plates, \citet{berthold83}\footnote{\url{https://www.sternwarte-hartha.de/wp-content/uploads/2018/11/Heft-18.pdf}} proposed V680 Mon to be an RR Lyrae star and derived first (but, in hindsight, incorrect) elements. Presumably based on this information, V680 Mon was included in the RR Lyrae star catalogues of \citet{mennessier02} and \citet{maintz05}. However, on the basis of ASAS-3 and NSVS observations, \citet{otero06} identified V680 Mon as an eclipsing binary star, in accordance with the initial proposition of \citet{parenago46}. The authors derived a period of $P$\,=\,8.5381\,d and a variability range of 9.93 $-$ 10.31\,mag ($V$). The system was found to be eccentric, with the secondary minimum occurring at phase $\varphi$\,=\,0.865. In consequence, V680 Mon entered the catalogue of eclipsing binary stars with eccentric orbits by \citet{bulut07} and new observations of minima were procured by \citet{brat09} and \citet{huebscher11}.

Despite these results, and the correct identification of the star as an eclipsing binary in the International Variable Star Index (VSX; \citealt{VSX}) of the American Association of Variable Star Observers (AAVSO), the star has been listed as an RR Lyrae variable in the SIMBAD database \citep{SIMBAD} until recently, which is probably the reason why it was included into the samples of the RR-Lyrae-star-based studies of \citet{gavrilchenko14} and \citet{gaia_parallaxes_17}. On the initiative of the present authors, V680 Mon is now correctly identified in the SIMBAD database as an eclipsing binary star.

\subsection{Sky region} \label{subsec:sky_region}

V680 Mon is situated in an area relatively devoid of bright stars, roughly in the midst of an imaginary triangle with Alhena ($\gamma$ Gem), Procyon ($\alpha$ CMi) and the Rosette Nebula (NGC 2244) as its vortices. However, situated at a distance of 12\arcsec\ from our target star, there is the relatively bright star 2MASS J06593015+0919070 = GAIA DR2 3157882744564424704 ($G$\,=\,12.52\,mag, \citealt{gaia2}). Both stars appear as a close double in DSS2 images (Fig. \ref{skyview}). They will also be blended in the spectroscopic and photometric data that form the backbone of this investigation; hence, a more detailed investigation into this matter is necessary.

Parameters of both stars are given in Table \ref{table_companion}. 
\citet{2020arXiv201205220B} list distances of, respectively, 619\,pc (581$-$672\,pc) and 2556\,pc (2438$-$2649\,pc) for V680 Mon and 2MASS J06593015+0919070 based on the Gaia Early Data Release 3 \citep[EDR3,][]{2020arXiv201201533G}. The two stars are not physically connected to each other. Considering its colour and luminosity, 2MASS J06593015+0919070 is obviously a late G- or early K-type giant. It is about 10 times ($\sim$2.5\,mag) fainter than V680 Mon, and we expect no significant contamination of the here employed spectra, which do not show any traces of the signature of a late G- or early K-type giant. Furthermore, we can rule out that the eclipses observed in the combined light curve of both stars originate in 2MASS J06593015+0919070. Even its complete disappearance would not result in a 25\,\% reduction in brightness, as is observed during primary eclipse of the system. We are therefore confident that no blending issues affect our main results.

\subsection{Observations} \label{subsec:observations}

The spectra employed in this study were extracted from the archive of the Large Sky Area Multi-Object Fiber Spectroscopic Telescope (LAMOST)\footnote{\url{http://www.lamost.org}}, which is located at Xinglong Observatory in Beijing (China), and procured at Star\'{a} Lesn\'{a} Observatory (SLO) and Skalnat\'{e} Pleso Observatory (SPO) in the High Tatras (Slovak Republic).

The LAMOST low-resolution spectrum boasts a resolution of R\,$\sim$\,1800 and covers the wavelength range from 3700 to 9000\,\AA. More information on the LAMOST survey is provided in \citet{lamost1} and \citet{lamost2}. The spectra taken at SLO have a resolution between R = 11\,000 and R = 12\,000 and cover the spectral range from 4150 to 7600\,\AA, while the spectra procured at SPO have a resolution of R = 38\,000 and cover the interval from 4250 to 7375\,\AA. More information on the instrumentation of the SLO and SPO observatories and the reduction process can be found in \citet{2015AN....336..682P}.

The photometric observations used in this study were procured by NASA's Transiting Exoplanet Survey Satellite (TESS), which provides ultra-precise photometry in a single passband (600-1000\,nm) taken at a cadence of 2\,min. More information on the TESS spacecraft and data products can be found in \citet{TESS3,TESS1,TESS2}.

\begin{figure*}
        \includegraphics[width=\textwidth]{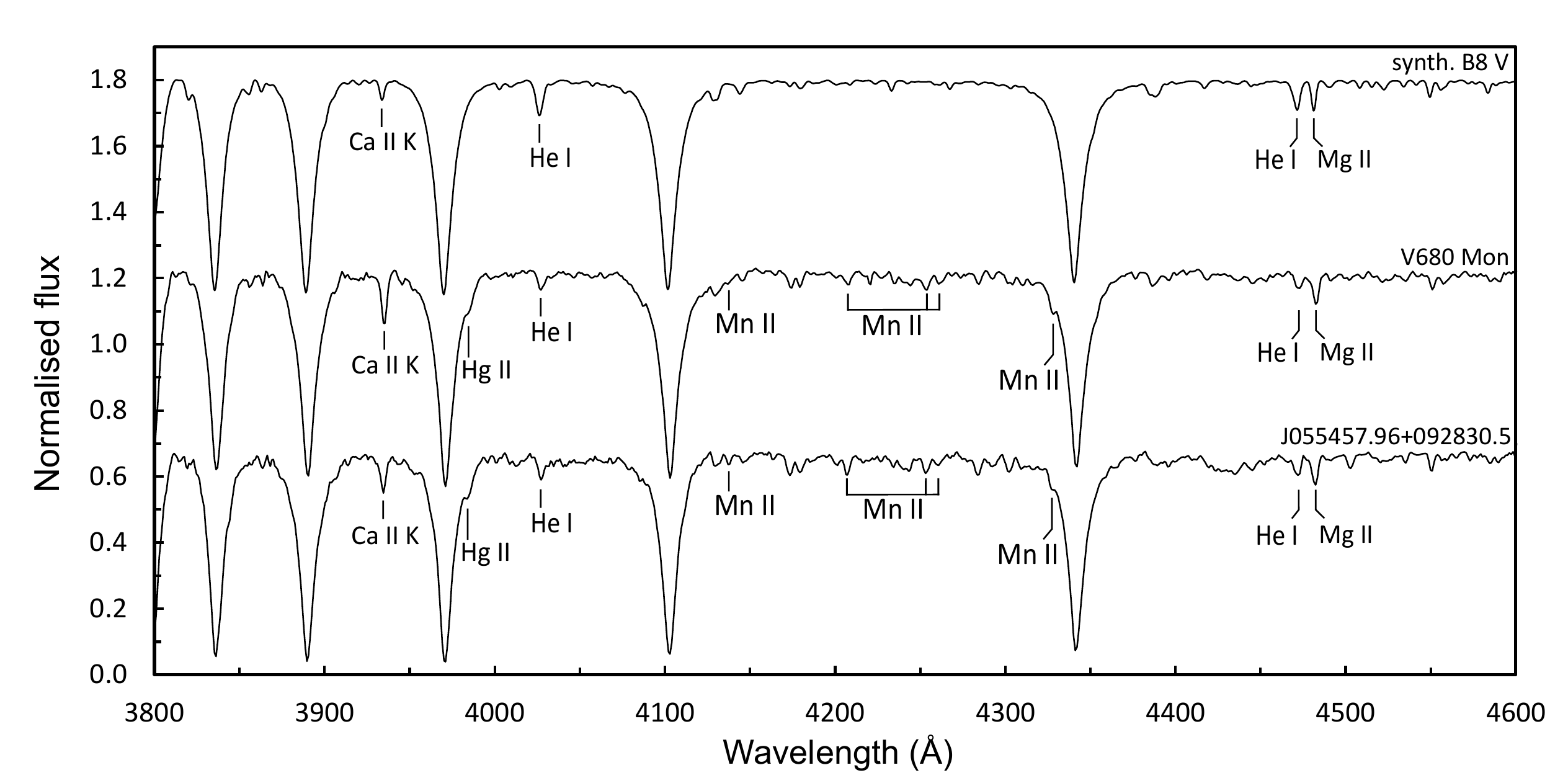}
    \caption{Comparison of the blue-violet region of (a) a synthetic spectrum corresponding to spectral type B8 V (${T}_{\rm eff}$\,=\,12500\,K; $\log g$\,=\,4.0; [M/H]\,=\,0.0; $\xi$\,=\,2\,km\,s$^{-1}$; smoothed to match the LAMOST resolution), (b) the LAMOST DR4 spectrum of V680 Mon (kB9 hB8 HeB9 V HgMn), and (c) the LAMOST DR4 spectrum of the CP3 star HD 249170 = LAMOST J055457.96+092830.5  (B8 IV HgMn; \citealt{paunzen20}). Some prominent lines of interest are identified.}
    \label{showcase1}
\end{figure*}

\section{Results} \label{sec:results}

\subsection{Spectral classification} \label{subsec:SpT}

The spectral peculiarities of V680 Mon were discovered during our semi-automated search for new CP3 stars in the spectra from LAMOST DR4 \citep{paunzen20}. Because it was recognised as an eclipsing binary and further studies were deemed necessary, V680 Mon was not included into the sample of \citet{paunzen20}. More information on our search for CP stars with Richard O. Gray's MKCLASS code, a program that classifies spectra by emulating the workflow of a human classifier \citep{gray14}, can be found in \citet{huemmerich20} and \citet{paunzen20}.

Only one LAMOST spectrum is available for V680 Mon = LAMOST J065930.88+091859.6, which was accessed via the DR4 VizieR online catalogue\footnote{\url{http://cdsarc.u-strasbg.fr/viz-bin/cat/V/153}} \citep{DR4}. The spectrum was obtained on 28 December 2015 (MJD 57384; observation median UTC 17:38:00; $g$-band S/N: 294), that is, at an orbital phase of $\varphi$\,=\,0.705. It was therefore obtained during maximum light and is dominated by light from the primary component of the V680 Mon system. The spectrum is illustrated in Fig. \ref{showcase1}, together with the synthetic spectrum of a B8 V star and the LAMOST DR4 spectrum of an HgMn star from the list of \citet{paunzen20}. 

Following the workflow outlined in \citet{paunzen20}, V680 Mon is given a final classification of B9 III$-$IV HgMn by the employed specialised version of the MKCLASS code. The \ion{Ca}{ii} K line strength and the relatively weak \ion{He}{ii} lines indeed suggest a spectral type of B9; the hydrogen line profile, however, is best matched by that of a B8 V standard. In general, the hydrogen-line profile is the most reliable indicator of the effective temperature in a CP star \citep{gray09}. Furthermore, CP3 stars have been shown to exhibit a large spread of He abundances, with most (usually the hotter) CP3 stars being He deficient \citep{smith96,ghazaryan16}. We therefore prefer a temperature type of B8 V. While CP3 stars also show a large dispersion of Ca abundances (up to 3 dex; \citealt{ghazaryan16}), we suspect that there is an interstellar contribution to the strong \ion{Ca}{ii} K line in the available spectrum.

The CP3 star characteristics are clearly present in the spectrum of V680 Mon. The \ion{Hg}{ii} 3984\,\AA\ line appears merely as a 'bump' in the red wing of H$\epsilon$, as is commonly the case at this low resolution (cf. e.g. the CP3 star spectra shown in \citealt{paunzen20}). The \ion{Mn}{ii} features at 4136\,\AA\ and, in particular, 4206\,\AA\ and 4152/9\,\AA\ are well developed (Fig. \ref{showcase1}). In summary, following the refined classification system of \citet{garrison94}, we arrive at a final classification of kB9 hB8 HeB9 V HgMn.

We analysed the high-resolution SLO and SPO spectra obtained at five different orbital phases and found no traces of the secondary component. V680 Mon, therefore, is a single-line spectroscopic binary (SB1) system. In addition to the classical classification criteria discussed above, we measured the equivalent widths of the \ion{Mn}{i} lines at 4462.031, 4762.367, 4765.846, 4766.418, 4783.427, 4823.524, and 6021.790\,\AA, all of which exhibit values between 15 and 20\,m\AA. This is well in line with the results obtained for the HgMn star HD 175640 ([Mn]\,=\,+2.45\,dex and [Hg]\,=\,+4.72\,dex as compared to the Sun; \citealt{castelli04}), which has an identical effective temperature as our target star. We also investigated the weak \ion{Hg}{i} 5460.731\,\AA\ line, which yields an equivalent width of 3\,m\AA. While this is at the detection limit of our set of spectra, its presence clearly indicates a significant overabundance of Hg; for solar metallicity, the equivalent width of this line is well below 1\,m\AA. In summary, our analysis of the SLO and SPO spectra corroborates the results from the blue-violet spectral region and confirms that the primary component of V680 Mon is a CP3 star.

\begin{figure*}
        \includegraphics[trim = 20mm 91mm 20mm 105mm, clip, width=18cm]{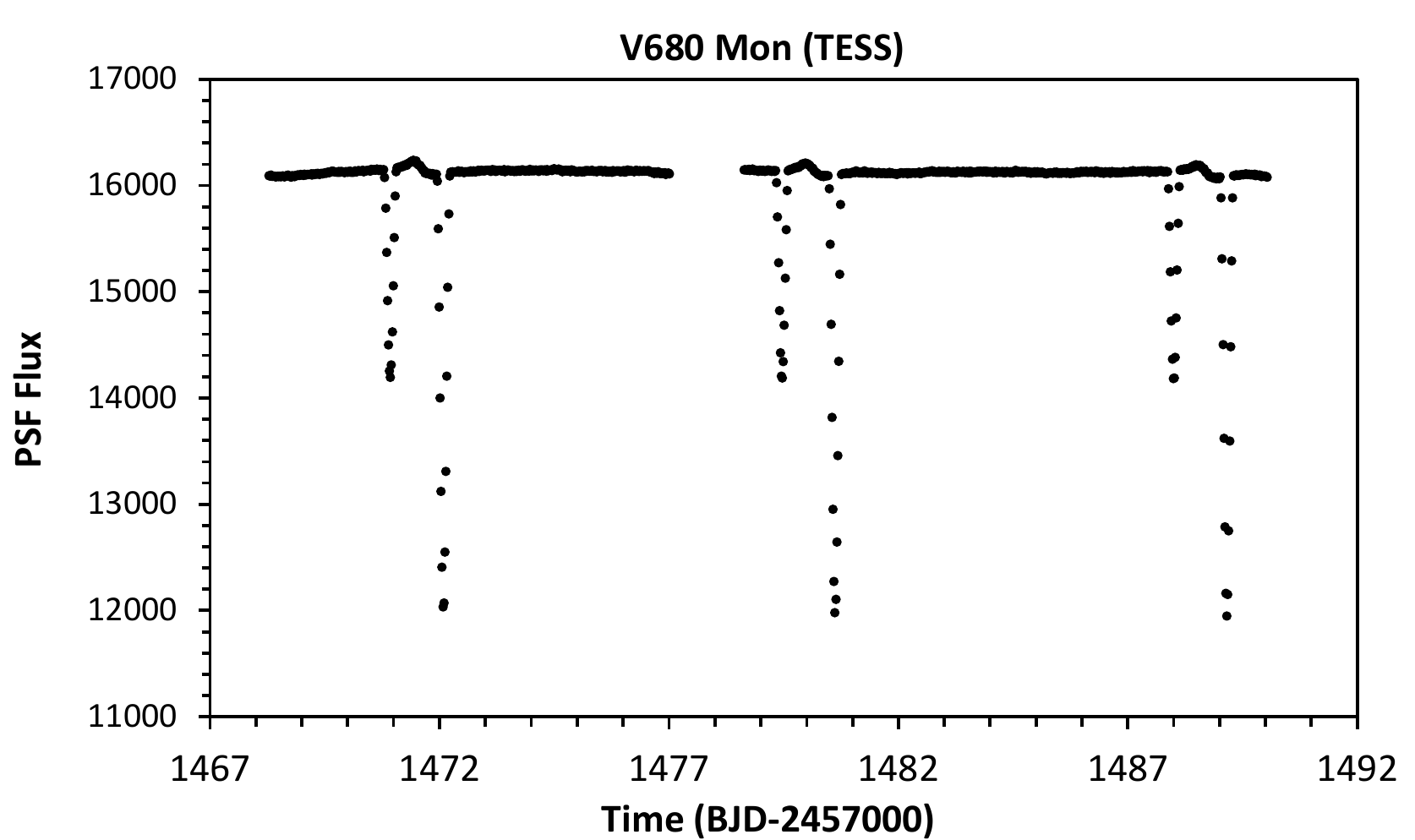}
    \caption{TESS light curve of V680 Mon, accessed via 'eleanor' and based on PSF flux.}
    \label{LCTESS1}
\end{figure*}

\begin{figure*}
        \includegraphics[trim = 20mm 107mm 26mm 120mm, clip, width=17cm]{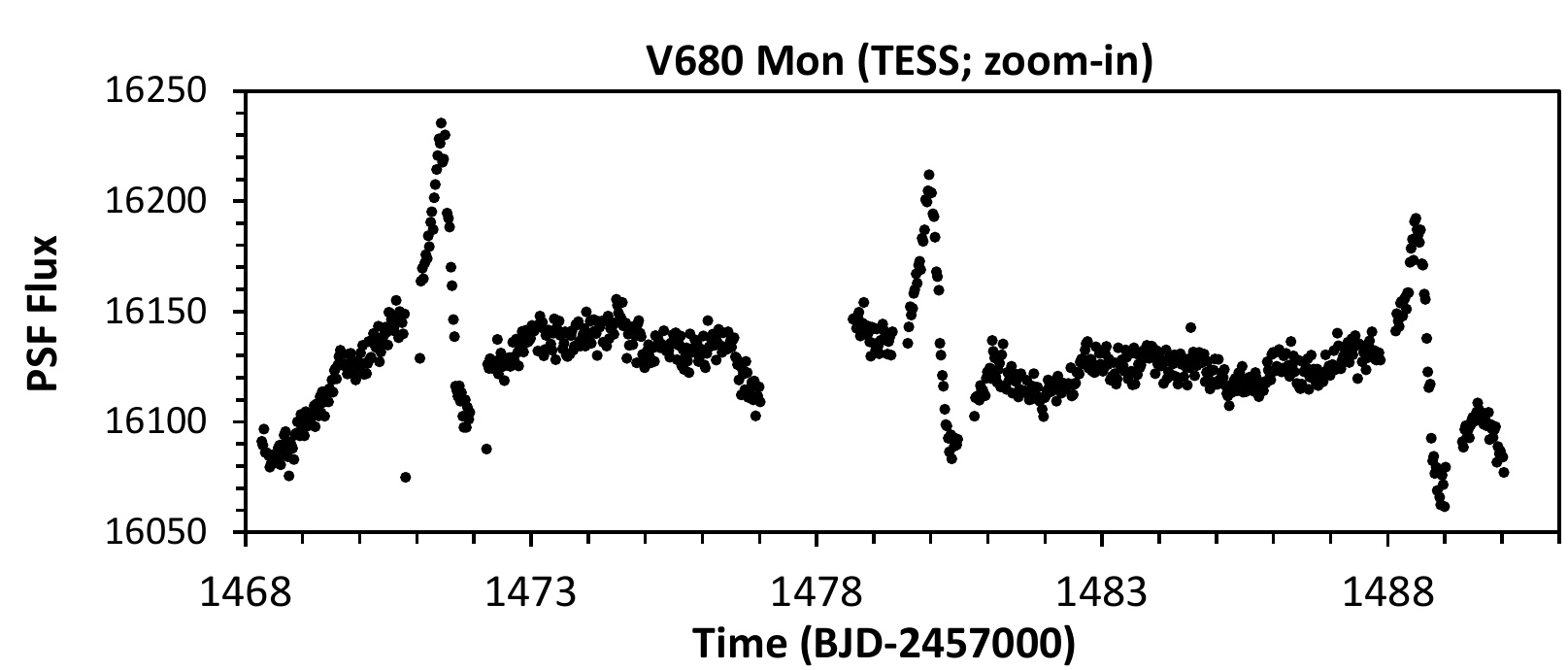}
    \caption{TESS light curve of V680 Mon, accessed via 'eleanor' and based on PSF flux, presenting a detailed view of the 'heartbeats'.}
    \label{LCTESS2}
\end{figure*}

\subsection{Photometric variability} \label{subsec:photvar}

On the basis of ASAS-3 and NSVS observations, \citet{otero06} identified V680 Mon as an eclipsing binary star (cf. Section \ref{subsec:target_star}). They derived a period of $P$\,=\,8.5381\,d (epoch of primary miminum: HJD 2452990.717), a total range of light variability of 9.93\,$-$\,10.31\,mag ($V$) and found the system to be eccentric, with the secondary minimum occurring at phase $\varphi$\,=\,0.865. With these data, the star was included into the VSX.

V680 Mon was observed by TESS during orbits 19 and 20 (TESS Observation Sector 6). The corresponding data were accessed via 'eleanor', which is an open-source python framework for downloading, analysing, and visualising data from the TESS Full Frame Images \citep{feinstein19}.\footnote{\url{https://adina.feinste.in/eleanor/}} In the direct vicinity of our target star, there is the relatively bright star 2MASS J06593015+0919070 ($G$\,=\,12.52\,mag), which is separated from V680 Mon by a distance of approximately 12\arcsec\ (cf. Section \ref{subsec:sky_region}). Since the TESS pixel size is relatively large (21\arcsec), the TESS light curve is a blending of light from both stars. However, as discussed in Section \ref{subsec:sky_region}, the light contribution from 2MASS J06593015+0919070 is negligible and cannot account for the observed eclipses. Apart from its close neighbor, V680 Mon is rather isolated from other bright stars. We tried both the PCA and PSF reductions and found that the PSF-modelled light curve is superior. For the final analysis, PSF modelling using a field of 7x7 pixels was employed.

The TESS light curve of V680 Mon is illustrated in Fig. \ref{LCTESS1}. Although the covered time-span is short, the ultra-precise TESS data reveal for the first time that V680 Mon is a heartbeat system. A detailed view of the heartbeats near periastron, whose shape is due to the combined effects of tidal distortion, reflection and Doppler beaming \citep{hambleton13,fuller17,hambleton18}, is presented in Figure \ref{LCTESS2}. The presence of chemical peculiarities in the component of a binary of 'heartbeat configurating' is an interesting find that is further discussed in Section \ref{sec:discussion}.

To derive an updated ephemeris, we combined the available photometric time-series data from TESS, the All Sky Automated Survey (ASAS-3; \citealt{ASAS1}) and the Kamogata/Kiso/Kyoto wide-field survey (KWS; \citealt{KWS}) and derived the elements presented in Eqs. \ref{eq1} and \ref{eq2}.

\begin{equation}
Min\,I = HJD\,2458472.088(1) + 8.53797(2)\,E \\
\label{eq1}
\end{equation}

\begin{equation}
Min\,II = HJD\,2458470.929(2) + 8.53797(2)\,E
\label{eq2}
\end{equation}

\begin{table}
    \caption{Parameters of the V680 Mon system as obtained from the Least-Squares Trust Region Reflective Algorithm ('Least Squares') and subsequent error estimation using the Markov Chain Monte Carlo (MCMC) sampler. The components' radii are given in semi-major axis units (SMA).}
    \label{table:EB_params}
    \centering
        \begin{tabular}{l c c c}
        \hline\hline
        Parameter                                      & \multicolumn{2}{c}{Value}                                          & Status      \\
        \hline                                                                                                              
        System                                         &             &                                                      &             \\ 
        \hline                                                                                                              
        $q$                   & \multicolumn{2}{c}{$0.57_{-0.02}^{+0.01}$}                         & Variable    \\
        $i[^{\circ}]$         & \multicolumn{2}{c}{$85.71_{-0.08}^{+0.08}$}                        & Variable    \\
        $e$                   & \multicolumn{2}{c}{$0.6131_{-1\times10^{-4}}^{+1\times10^{-4}}$}   & Variable    \\
        $\omega[^{\circ}]$    & \multicolumn{2}{c}{$356.36_{-0.08}^{+0.09}$}                       & Variable    \\
        $P[d]$                & \multicolumn{2}{c}{$8.53797$}                                      & Fixed       \\
        $T_0[d]$              & \multicolumn{2}{c}{$2458472.088$}                                  & Fixed       \\
        \hline
        \hline
        Component             & primary                       & secondary                          &             \\ 
        \hline
        $\Omega$       & $12.62_{-0.04}^{+0.04}$       & $13.2_{-0.2}^{+0.2}$               & Variable    \\
        $F$            & $5.1_{-0.4}^{+0.5}$           & $5.1_{-0.2}^{+0.3}$                & Variable    \\
        $r_{eq}$       & $0.0836_{-0.0002}^{+0.0002}$  & $0.0474_{-0.004}^{+0.005}$         & Derived     \\
        \hline
        \multicolumn{4}{l}{Atmospheric parameters}                                  \\
        \hline
        $T^{eff}/[K]$   & $12000_{-300}^{+400}$         & $8300_{-200}^{+200}$               & Variable    \\
        $\beta$         & $0.86_{-0.05}^{+0.04}$        & $0.89_{-0.04}^{+0.03}$             & Variable       \\
        $A$      & $0.73_{-0.07}^{+0.07}$        & $0.64_{-0.04}^{+0.04}$             & Variable       \\
        $M/H$           & 0.0                           & 0.0                                & Fixed       \\
        \hline
        \multicolumn{4}{l}{Radii at periastron} \\
        \hline
        $r_{pole}$      & $0.0827_{-0.0003}^{+0.0003}$  & $0.0472_{-0.0004}^{+0.0005}$       & Derived     \\
        $r_{back}$      & $0.0844_{-0.0002}^{+0.0002}$  & $0.0476_{-0.0004}^{+0.0005}$       & Derived     \\
        $r_{side}$      & $0.0837_{-0.0002}^{+0.0002}$  & $0.0474_{-0.0004}^{+0.0005}$       & Derived     \\
        $r_{forward}$   & $0.0846_{-0.0002}^{+0.0002}$  & $0.0476_{-0.0004}^{+0.0005}$       & Derived     \\
        \hline\hline
        \end{tabular}
\end{table}

\begin{figure*}
 \includegraphics[width=1.0\linewidth]{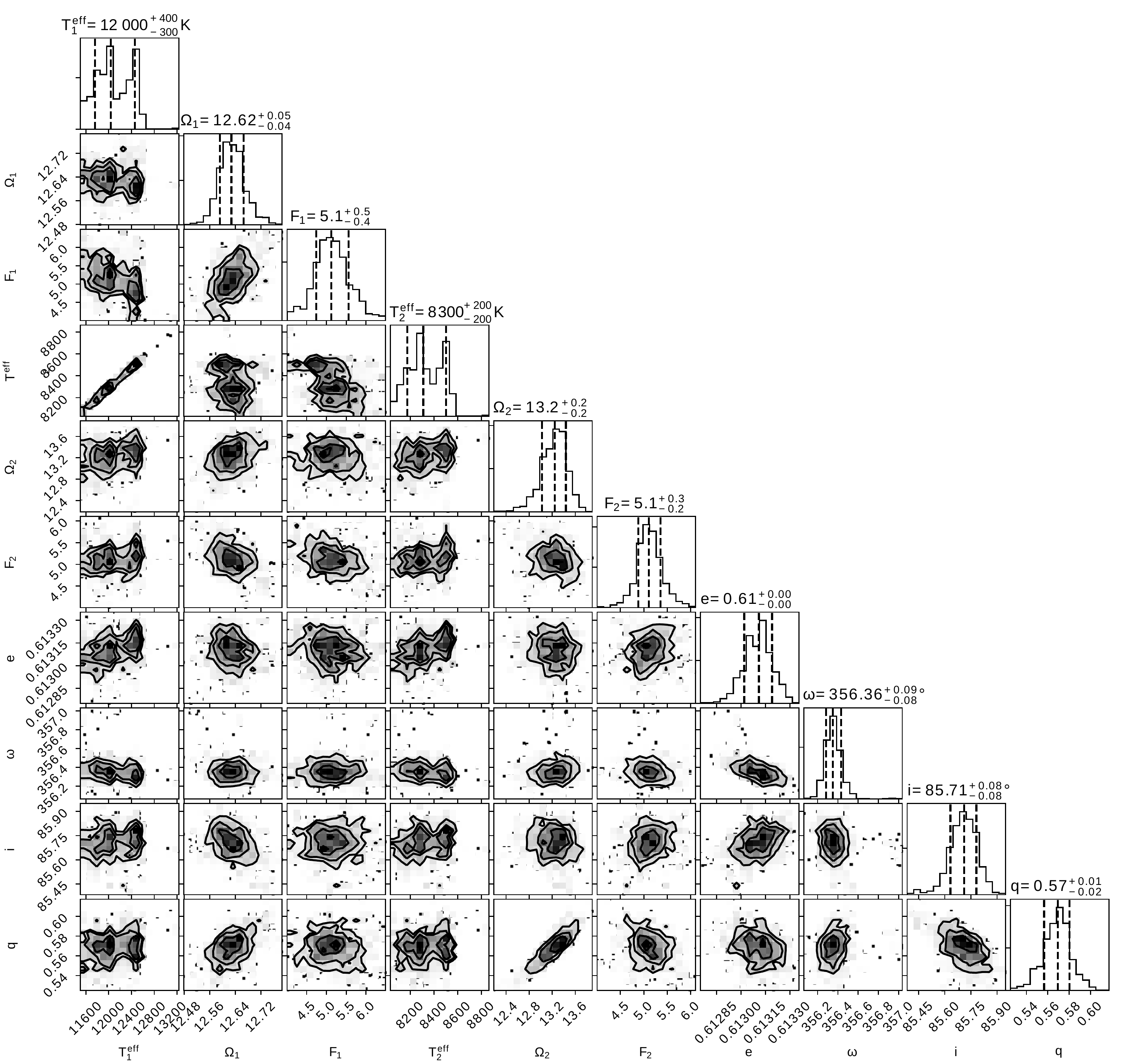}
 \caption{Posterior distribution of the samples generated by the Markov Chain Monte Carlo (MCMC) algorithm. \label{fig:corner_plot}}
\end{figure*}

\subsection{Binary system modelling} \label{subsec:modelling}

Due to the significantly lower quality of the ASAS-3 and KWS photometry, binary system parameters were inferred using only detrended TESS photometry. The light curve was analysed with the Python package ELISa\footnote{\url{https://github.com/mikecokina/elisa}}, which contains tools for modelling light curves of close eclipsing binaries utilising Roche geometry and methods for solving an inverse problem.

As a first step, the Least-Squares Trust Region Reflective Algorithm ('Least Squares' hereafter) was used to search for a local minimum around a manually-selected starting point based on the general shape of the light curve. Initial runs were performed with seven free parameters: photometric mass ratio $q$; inclination $i$; eccentricity $e$; argument of periastron $\omega$; surface potentials $\Omega_1$, $\Omega_2$; and effective temperature of the secondary component $T^{eff}_2$. The effective temperature of the primary component $T^{eff}_1$ was fixed to 12\,500\,K, as inferred from the spectral classification (cf. Section \ref{subsec:SpT}). Albedos $A_1$, $A_2$ and gravity darkening factors $\beta_1$, $\beta_2$ were set to 1.0 since both components were expected to have radiative envelopes due to their effective temperatures being well above 7000\,K \citep{zeipel24}. We adopted a square-root law for limb darkening. The corresponding coefficients were interpolated from the pre-calculated tables of \citet{Claret17}, and atmospheric models from \citet{castelli04b} were used for the calculation of the integrated passband flux. Finally, the synchronicity factors $F_1$, $F_2$ were set to assume synchronous rotation of the components at periastron \citep{Hut81}.
 
After the initial solution was found, the effective temperature of the primary component $T^{eff}_1$, synchronicity parameters $F$, gravity darkening factors $\beta$, and albedos $A$ were set variable to allow for the model to relax into the local minimum. $T^{eff}_1$ was allowed to vary $\pm$1000\,K. Using the approach described above, the Least Squares algorithm arrived at a solution with a coefficient of determination of $R^2 = 0.9990$.

\begin{figure*}
 \includegraphics[trim = 0mm 0mm 8mm 140mm, clip, width=\textwidth]{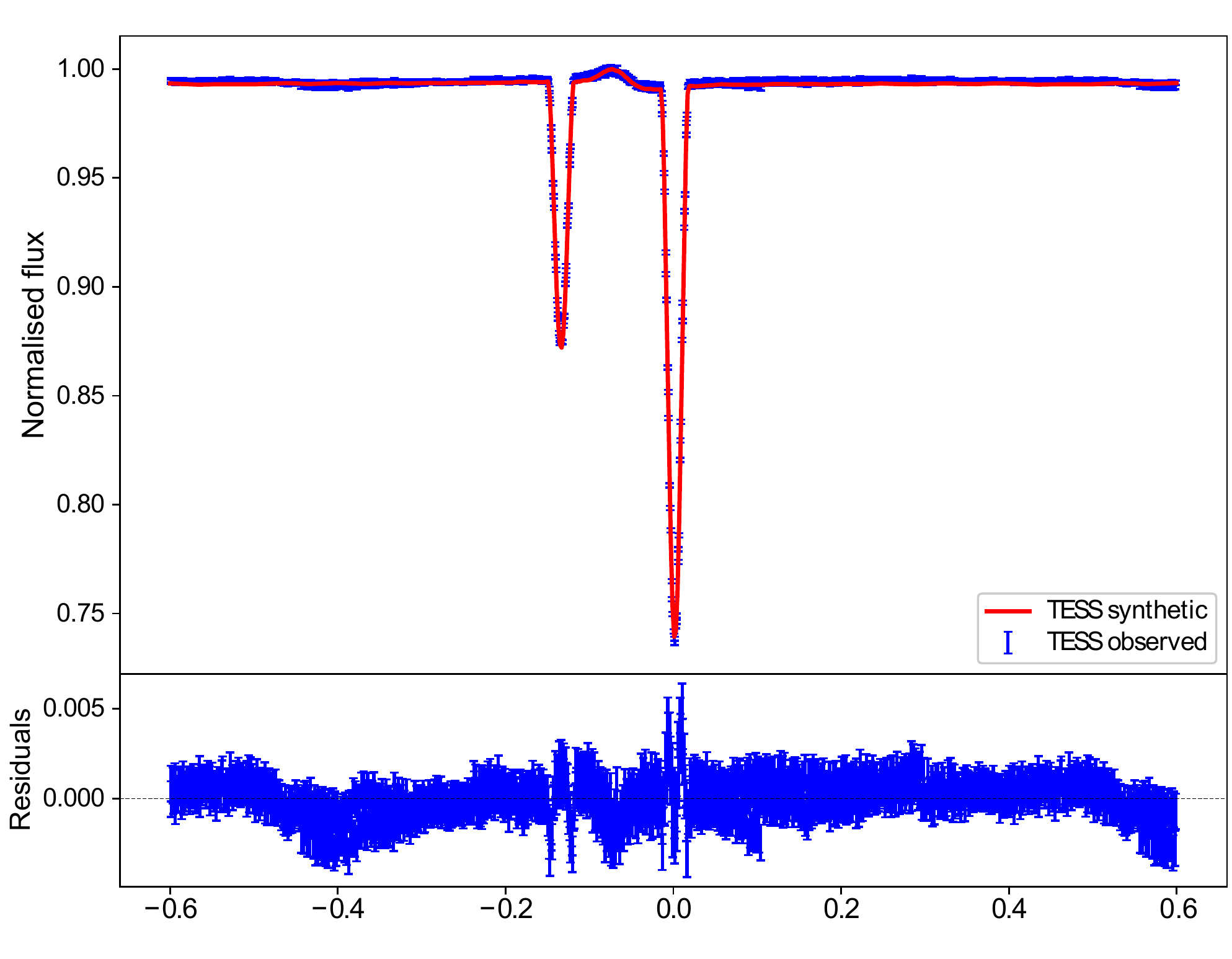}
 \caption{The upper panel illustrates the fit between observed and synthetic flux in the TESS passband based on the best-fit model presented in Table \ref{table:EB_params}. The fitting residuals are shown in the lower panel.}
 \label{fig:lc_fit}
\end{figure*}

The vicinity of the obtained solution was sampled using the Markov Chain Monte Carlo algorithm (MCMC) with 200 walkers, 200 steps, and uniform prior sampling. After discarding the thermalisation phase of the chain, the confidence intervals of the parameters were inferred from the posterior distribution displayed in Figure \ref{fig:corner_plot} in the form of a corner plot. The resulting system parameters are listed in Table \ref{table:EB_params} and the corresponding fit is illustrated in Figure \ref{fig:lc_fit}. 

The achieved solution points to an eccentric orbit with eccentricity $e=0.61$ and a relatively short orbital period. The tidal forces during the periastron passage result in a deformation of both components with an amplitude of $\sim$1\,\% (ratio between forward and equivalent radius). The synchronicity parameters for both components remained within errors around the predicted value of $5.27$, which suggests that the rotation of the components is synchronised with the orbital motion during the periastron. Figure \ref{fig:corner_plot} shows the double Gaussian distribution for both effective temperatures that indicates two solutions with similar qualities of fit. However, since their spacing is similar to the standard deviations of each peak, we decided to regard them as a single solution. No additional double Gaussian distributions were detected for any other variable parameters. We note that the fitting process based on information obtained in a single passband has a limited capacity to recover information about the effective temperature of the components.

\section{Discussion} \label{sec:discussion}

The evolutionary stages of the four known eclipsing binary systems containing a CP3 star component are widely different \citep{gonzalez14,kochukhov20},
ranging from close to the zero-age main sequence (HD 34364 and TYC 455-791-1)
to the middle of the main sequence (HD 161701) and close to the terminal-age main sequence (HD 10260).

To evaluate the evolutionary status of V680 Mon, we investigated its position in the Hertzprung-Russell diagram. As first step, the absolute magnitude listed in Table \ref{table_companion} was corrected for
the contribution of the secondary component as shown in Fig. 9 of \citet{paunzen20}. For the given $q$ value (Table \ref{table:EB_params}), this correction amounts to 0.15\,mag only. The bolometric correction for CP stars \citep{2008A&A...491..545N} yields a value of $-$0.559\,mag and thus a luminosity of $\log L/L_\odot$\,=\,1.628(64). For the calibration of the age, we used the Stellar Isochrone Fitting Tool\footnote{\url{https://github.com/Johaney-s/StIFT}} and the isochrone grid by \citet{2012MNRAS.427..127B} for solar metallicity.

Our results indicate that V680 Mon is located on the zero-age main sequence with an age between 5 and 6\,Myr. This result is not dependent on the choice of the isochrone metallicity because the use of overabundant isochrones would result in an even larger luminosity for the same effective temperature. In such grids, our target star would be situated significantly below the zero-age main sequence.

Several CP3 stars belong to open clusters \citep[e.g.][]{hubrig12}, which puts further constraints on the age determination. We searched for a possible host cluster within 3$\sigma$ of the position, diameter, proper motion, distance and their errors of the star cluster lists from \citet{Dias2002} and \citet{Cantat2020}. Because V680 Mon is quite close (about 620\,pc from the Sun), we expect these lists to be complete. The closest aggregate is NGC\,2264, which is located about 4.6 degrees or 50\,pc away. In this young open cluster, star formation is still going on \citep{2021A&A...645A..94N} and many young stellar objects are present \citep{2020A&A...636A..80B}. If we accept V680 Mon as a member of NGC\,2264, the derived ages are in excellent agreement. Incidentally, another HgMn star, HD 47553, was reported as a member of NGC\,2264 \citep{1993BSAO...35...76P}; unfortunately, besides this reference, no other analysis of this star was found in the literature.

V680 Mon is only the fifth known eclipsing CP3 star, and it is the first one recognised as member of a heartbeat binary. Our results indicate that the V680 Mon system is composed of a CP3 star primary component ($T_{eff}$\,=\,$12000_{-300}^{+400}$\,K; spectral type kB9 hB8 HeB9 V HgMn) and a secondary component of spectral type A4 ($T_{eff}$\,=\,$8300_{-200}^{+200}$; cf. Section \ref{subsec:modelling}). The unique combination of a very young and relatively bright chemically peculiar star in such a system opens up intriguing possibilities. In particular, theory needs to explain the development of CP3 star features in such a young object and under the conditions (tidally-induced effects) of a heartbeat binary. Our modelling attempts indicate a significant deformation of both components by about one per cent (ratio between forward and equivalent radius) during periastron passage.

In this respect, it is interesting to note that there is a high rate of occurrence of very eccentric short-period binary systems among the CP1 stars \citep{debernardi00}, which some studies have proposed as lower-temperature counterparts of the CP3 stars \citep{adelman03}. Obviously, the tidally-induced effects do not interfere with the development of CP1 star abundance patterns in these systems. CP2, CP3 and CP4 stars, on the other hand, show a rather different eccentricity versus orbital period distribution with an apparent upper envelope \citep{carrier02}. Interestingly, V680 Mon is located well above this proposed upper envelope (cf. Fig. 8 of \citealt{carrier02}).

In summary, V680 Mon lends itself perfectly for detailed follow-up studies and may prove to be a keystone in the understanding of the development of CP3 star peculiarities.

\section*{Acknowledgements}
We thank the referee, Gautier Mathys, for his comments that helped to improve the paper. We furthermore thank Theodor Pribulla and Johana Sup{\'i}kov{\'a} for their help in preparing this manuscript and express our gratitude to Iosif I. Romanyuk for providing the original Peremennye Zvezdy paper of P. P. Parenago and Hans Michael Maitzen for his translation of the Russian original. EP acknowledges support by the Erasmus+ programme of the European Union under grant number 2020-1-CZ01-KA203-078200. This work has been supported by the VEGA grants of the Slovak Academy of Sciences No. 2/0031/18 and 2/0004/20. This paper makes use of data from the Guoshoujing Telescope (the Large Sky Area Multi-Object Fiber Spectroscopic Telescope LAMOST), which is a National Major Scientific Project built by the Chinese Academy of Sciences. Funding for the project has been provided by the National Development and Reform Commission. LAMOST is operated and managed by the National Astronomical Observatories, Chinese Academy of Sciences. Furthermore, this paper includes data collected by the TESS mission. Funding for the TESS mission is provided by the NASA's Science Mission Directorate.
This work presents results from the European Space Agency (ESA) space mission Gaia. Gaia data are being processed by the Gaia Data Processing and Analysis Consortium (DPAC). Funding for the DPAC is provided by national institutions, in particular the institutions participating in the Gaia MultiLateral Agreement (MLA). The Gaia mission website is https://www.cosmos.esa.int/gaia. The Gaia archive website is https://archives.esac.esa.int/gaia.
This research has made use of the WEBDA database, operated at the Department of Theoretical Physics and Astrophysics of the Masaryk University, and the SIMBAD and VizieR databases, operated at CDS, Strasbourg, France.

\section*{Data Availability}
The data underlying this article will be shared on reasonable request to the corresponding author.

\bibliographystyle{mnras}
\bibliography{V680Mon}

\bsp	
\label{lastpage}
\end{document}